\documentclass[prd,eqsecnum,noshowpacs,nofootinbib,notitlepage]{revtex4-1}
\usepackage{amsmath,amssymb,amsthm}
\usepackage{amsfonts}
\usepackage{comment}
\usepackage{color}
\usepackage[pdftex]{graphicx}
\usepackage{mathptmx}


\newtheorem{theorem}{Theorem}
\newtheorem*{theorem*}{Theorem}
\newtheorem{definition}{Definition}

\newtheorem{corollary}{Corollary}

\newtheorem{proposition}{Proposition}

\begin{document}
\title{Black hole shadow and Maximal Black room}

\author{Masaru Siino}
\email{msiino@th.phys.titech.ac.jp}
\affiliation{Department of Physics, Tokyo Institute of Technology, Tokyo 152-8551, Japan}

 \begin{abstract}
To comprehend the shadow of a black hole\cite{Falcke:1999pj,Broderick:2009ph,Broderick:2010kx} in a general spacetime, we have investigated the concept of the maximal black room (MBR). The boundary of the MBR is a non-spacelike hypersurface that contains at least one null geodesic tangent to its boundary, which we refer to as the rays' surface. Our aim is to explore the geometry of this surface.

From our current study, we have observed that the boundary of the MBR is globally unstable. Leveraging this instability, we can determine the boundary based on the physical choice of the initial spacelike hypersurface and the orthogonal condition at that point.

Upon examining the existence of the MBR, we can observe that it encompasses any arbitrary black hole. Additionally, we can investigate the possibility of using the MBR to differentiate a black hole from an exotic star with a photon sphere. In spherically symmetric spacetime, subject to energy conditions, the rays' surface encloses a black hole, a naked singularity, or exotic matter that has an inner universe.

 \end{abstract}

\maketitle
\section{Introduction}
The observation of a photon originating from the region in close proximity to a black hole has been reported\cite{Akiyama:2019cqa,Akiyama:2019brx,Akiyama:2019sww,Akiyama:2019bqs,Akiyama:2019fyp,Akiyama:2019eap}\cite{Wielgus2020}.
The image appears to provide information regarding both the geometry of the black hole and the distribution of the luminous sources surrounding it.
To comprehend the image, one may employ ray tracing\cite{NouriZonoz2022}, assuming the geometry and the distribution of the photon source\cite{Narayan2019,Gralla2019}.
As such, crucial evidence of spacetime geometry, particularly with regard to the existence of the event horizon, may not always be obtainable.
Furthermore, attempting to identify modifications to gravitation based on the image\cite{Zeng2022} constitutes an inconvenient and ill-posed inverse problem, lacking in both uniqueness and stability.
 
Roughly stated, the image can be divided into a luminous annular area and a dim central region. The annulus can be easily replicated through the application of Schwarzschild geometry, given the broad assortment of photon source distributions. 
Conversely, if one endeavors to derive data regarding the spacetime geometry utilizing a non-Schwarzschild spacetime assumption, the typical methodology involves assessing a distorted photon sphere\cite{Schneider2018,Cao2021,Mishra2019,Bisnovatyi-Kogan2018, Mars2017}, albeit the relationship between the photon surface\cite{Claudel:2000yi}\cite{Cao2021} and the image is generally indeterminate. How might we tackle this conundrum?

A robust methodology would entail generating a substantial quantity of ray tracing outcomes. Alternatively, an audacious endeavor would be to elucidate the connection between a comprehensive photon surface\cite{Claudel:2000yi} and the perceived image. For instance, the photon surface was posited as an extension of the photon sphere, as it aligns with the photon sphere in spherically static spacetime\cite{photonsphere,Hod2013,Hod2018}.

Focusing on the central dark region\cite{Sanchez1978,Decanini2010} may help clarify the problem at hand. By considering the maximal black room\cite{Siino2022} as a mathematical definition of the dark region, we observe that its boundary may not always be the photon surface, but rather a hypersurface, on which at least one direction of a light ray lies. In this article, we refer to this hypersurface as the `rays' surface'. While the bright ring is influenced by various complex astrophysical effects, the geometry may largely determine the central black region.

Thus, although the photon sphere could explain the image of the black disk in a Schwarzschild black hole, how can we extend this discussion beyond the spherically symmetric spacetime? To expand on the concepts of the photon sphere and related ideas, one approach is to examine them from the viewpoint of the causal structure of the spacetime.

In Ref. \cite{Siino2020}, we have demonstrated the wandering null geodesic can play a role of the photon circular
orbit.
Then introducing the black room which recognize the central absence of the light rays, the role of the
wandering
null geodesics becomes clear, so that the boundary of the maximal black room is a rays' surface generated by
the
wandering null geodesic or horizon generators\cite{Siino2022}.
Of course, in a spherically symmetric spacetime, the rays' surface is just a photon surface and should be a
photon sphere in static situations.
Therefore, we had better take an interest in the rays' surface as well as photon surface.

The objective of this article is to investigate the geometric properties of the maximal black room (MBR) by utilizing insights from the rays' surface. In the case of a static and spherically symmetric spacetime, the structure of the shadow of a black hole\cite{Falcke:1999pj,Broderick:2009ph,Broderick:2010kx} is described by a photon sphere. Hence, we possess considerable knowledge regarding the connection between the photon sphere and compact objects\cite{Claudel:2000yi}\cite{Cao2021}. Conversely, by examining the rays' surface, we may uncover further insights into the interior of the black hole shadow in the future, as the rays' surface is anticipated to exist in more general scenarios\cite{Kobialko2020}.

In the upcoming section, we shall review the findings from Refs.\cite{Siino2020}\cite{Siino2022}, which provide definitions for concepts concerning the black room and wandering null geodesics.

In the third section, we will present a demonstration of the global instability of the MBR. Furthermore, using this instability, we will show how to determine the MBR in a black hole spacetime.

The fourth section will expound upon the existence of the MBR, examining its relationship with the exotic star and exploring the conditions for its existence.

The final section shall be devoted to the conclusion.

\section{causal concept of the black hole shadow}
In the aforementioned references \cite{Siino2020}, the causal concept of the black hole shadow was studied to analyze its geometric properties. Initially, a null geodesic was considered which did not enter the event horizon or escape to null infinity. Then, the notion of the circular photon orbit was extended.

Let $(M,g)$ denote a globally hyperbolic asymptotically flat spacetime. Owing to asymptotic flatness, the unphysical spacetime manifold $\overline{M}$ features a spatial infinity $i^0$ and a future null infinity ${\cal I}^+$, defined as the causal future of $i^0$. Besides, future null infinity ${\cal I}^+$ admits a null coordinate $u$ in the form of ${\cal I}^+\simeq {(0,\infty)}\times S^2\sim {(u,x^1,x^2)|u\in (0,\infty)}$.
Given global hyperbolicity, we obtain a family of Cauchy surfaces ${\cal C}[t]$ as a family of spacetime's time slices such that $M\simeq {\cal C}[\cdot]\times [0,\infty)$. Subsequently, the family of time slices is deemed complete since $\overline{M}\supset J^-({\cal I}^+)$, even in the presence of a timelike or null singularity inside the event horizon.
\begin{definition}[neutral null geodesic]
A null geodesic that originates from a point $p$ is dubbed a {\it neutral null geodesic} starting from $p$ if, for any sufficiently large $u$, there exists a sufficiently large $t$ such that it does not cross the event horizon and ${\cal C}[t]\cap J^-({\cal I}^+[u])$.
\end{definition}

However, in the presence of spacetime curvature, such a neutral null geodesic must be accompanied by an infinite number of conjugate points, rendering it an essential aspect of the null geodesic orbiting around the black hole. Thus, we introduce the concept of wandering null geodesic.
\begin{definition}[Wandering Null Geodesic\cite{Siino2020}]
A future (past) wandering null geodesic is a complete null geodesic with an infinite number of conjugate points in the future (past) direction. Additionally, a totally wandering null geodesic is defined as a complete null geodesic that possesses an infinite number of conjugate points both in the future and past directions.
\end{definition}

According to Theorem 9.3.11 of the textbook \cite{Wald:1984rg}, the existence of conjugate points can be attributed to a null geodesic staying within the chronological set of the starting point\cite{Hawking:1973uf}.
It is worth noting that Ref\cite{Siino2022} demonstrates that a wandering null geodesic can start from any point in spacetime outside of the black hole region. This corresponds to null geodesics that infinitely wind around the photon sphere in the Schwarzschild spacetime.
Therefore, it is more accurate to say that the circular photon orbit corresponds to the totally wandering null geodesic.

To elucidate the connection between the optical image in the vicinity of the black hole and the wandering null geodesic, we shall introduce a black room ${\cal R}i$ and a maximal black room ${\cal R}{max}$ surrounding the black hole.

Let us define the maximal black room as follows:

\begin{definition}[Maximal Black Room]
Let ${\cal R}_i\subset M$ be each arc-wise connected (spatially) bounded open subset, such that no null geodesic can escape from ${\cal R}_i$ after entering ${\cal R}i$. If there exists a ${\cal R}{max}\in { {\cal R}_i| i=1,2...}$ that encompasses all ${\cal R}_i$'s, we refer to it as a maximal black room.
\end{definition}

It is evident that in the presence of a black room, any light sources located outside of it would not contribute to the optical image of $\cal R$. Consequently, in the absence of a light source contained within $\cal R$, the region would appear dark in optical imaging. This phenomenon is analogous to the photon sphere of Schwarzschild spacetime.
In Schwarzschild spacetime, the maximal black room exists and its outer boundary is identified with the photon sphere. As long as there is no photon source in the maximal black room (MBR) or in the foreground, photons never approach the MBR from this direction.
In a realistic scenario, it is improbable for a photon source to be positioned around the photon sphere or the maximal black room. The light signal emanating from realistic light sources, such as the accretion disk, would accumulate on the boundary of the MBR before reaching us (refer to section 4 of the cited reference\cite{Siino2020}).

From the global reason, we should consider a spacetime region chronologically complete in future direction $\sim\{{\cal C}(t)| t\in (0,\infty)\}$.
Thus far, we have considered the occurrence of black hole formation at some point in time. In such a scenario, the formation of the black room is anticipated. Nevertheless, given that the formation of the black room entails the entry of all null geodesics that escape from the region prior to the formation, the only permissible black room is that which corresponds to the black hole region itself.

\begin{figure}[hbtp]
\begin{center}
\includegraphics[height=12cm]{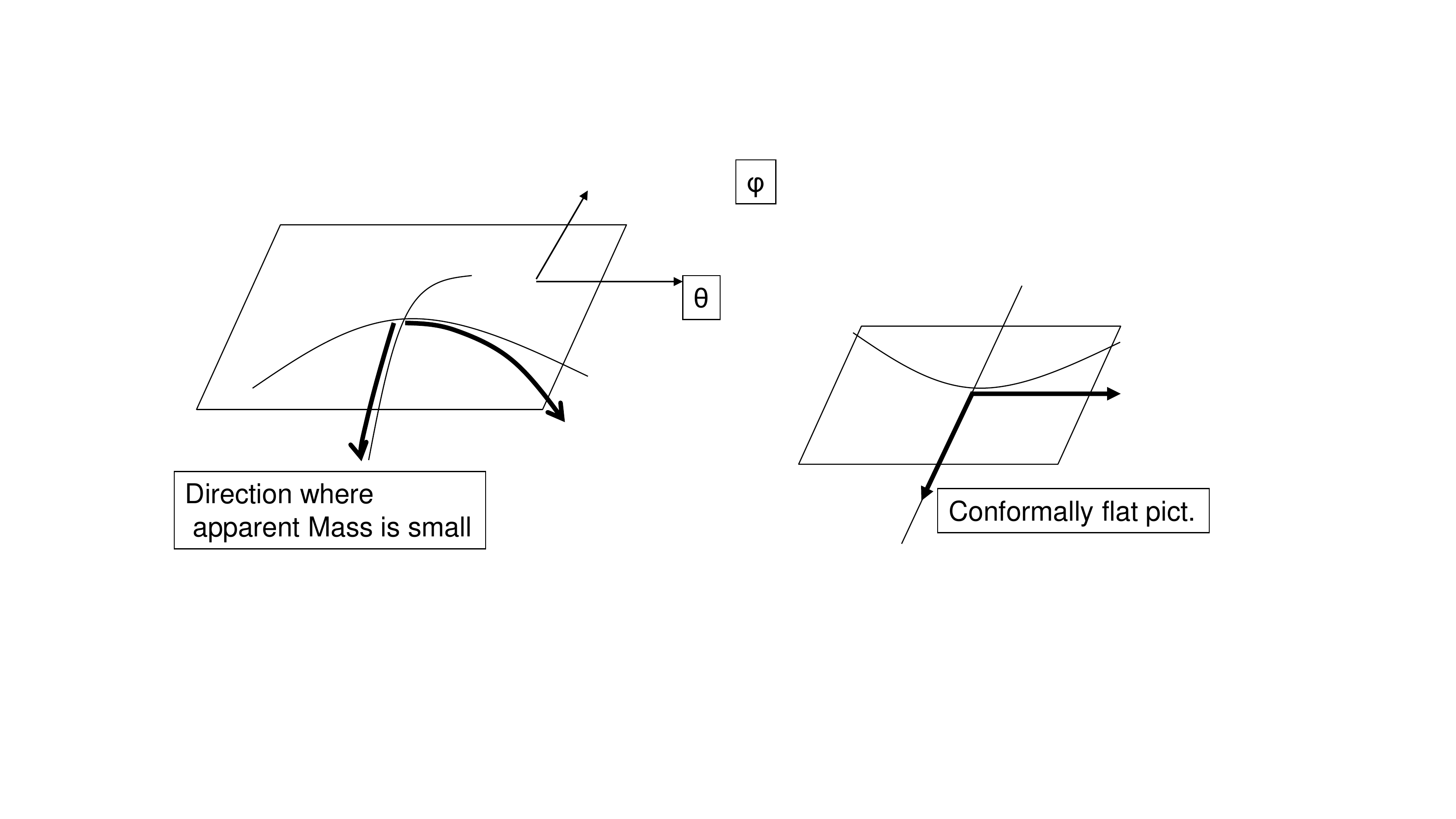}
\caption{At points located on the periphery of the maximal black room, there is at least one tangential null geodesic included in the boundary hypersurface. Denoting the extrinsic curvature as $K_{ab}$, the condition $K_{ab}k^ak^b\leq 0$ is satisfied, with equality achieved in the direction where the null geodesic lies. Hence, in the vicinity of this direction, the negative curved surface is found to be embedded with a minute degree of curvature, approaching zero. This phenomenon can be interpreted as the direction in which the black hole manifests the least apparent mass.}
 \label{fig:rays}
 \end{center}
\end{figure}

According to Proposition 2 of \cite{Siino2022}, the boundary of the MBR can be identified as a non-spacelike hypersurface embedded with non-positive extrinsic curvature. Specifically, denoting the extrinsic curvature of the boundary of the MBR in the outward direction as $K_{ab}$, we have $K^{ab}k^ak^b\leq 0$ for any null vectors $k^a$ that are tangential to $\partial{\cal R}$, and the equality is achieved at least in one direction (refer to Fig. \ref{fig:rays})\cite{Kobialko2020}. The equality can be achieved only when all the possible null geodesics are included in the surface, and it can be viewed as a photon surface\cite{Claudel:2000yi}.
If the surface is not a null surface, then a null geodesic that is wandering will be lying on it.
 \begin{figure}[hbtp]
\begin{center}
\includegraphics[height=12cm]{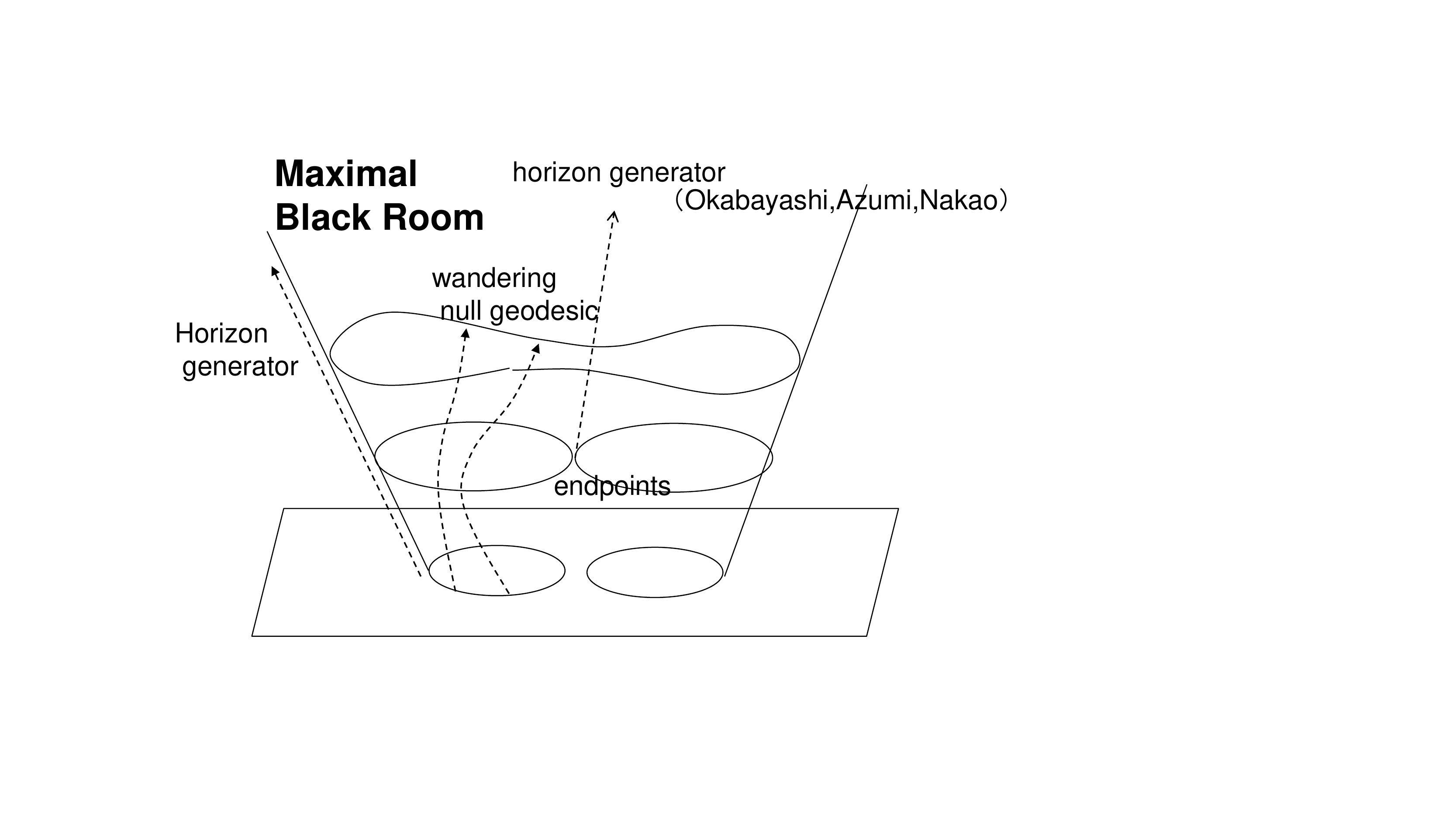}
\caption{The scenario at hand depicts a symmetric, head-on collision between two black holes of equal mass, both devoid of spin. Specifically, we are presented with the collision of two non-rotating black holes, which approach each other from opposite poles of the celestial sphere.
}
 \label{fig:col}
 \end{center}
\end{figure}
In Ref. \cite{Siino2022}, it has been elucidated that the MBR is bounded by a non-spacelike hypersurface that contains at least one tangential null geodesic at each point. In relation to this, we may conveniently define a rays' surface as an extension of the photon surface.

\begin{definition}[rays' surface]
The rays' surface is a non-spacelike hypersurface that is embedded with null components of extrinsic curvature, i.e., $K_{ab}k^ak^b\leq 0$, for every tangential null vector $k^a$. At each point of the surface, there is at least one tangential null geodesic encompassed within it.
\end{definition}

In this scenario, the majority of null geodesics would be considered as wandering null geodesics, except in the instance when the rays' surface is in contact with the event horizon. 
An example that illustrates the practical application of these concepts is when we examine an axisymmetric scenario involving colliding black holes.
To avoid the coincidence between the black hole and the MBR, suppose that there already two black hole have formed at the initial hypersurface.
If the MBR includes the collision of its spatially disconnected components, there must exist a past endpoint of the null geodesics contained within the boundary of the MBR. Conversely, the corresponding part of the boundary should be a null surface in order to have a past endpoint, as per \cite{Siino2022}. Therefore, only null generators of the event horizon are allowed there. This fact explains why an observer will never witness the merging of each component of the MBR, as stated in \cite{Okabayashi2020,Cunha2018}.

Furthermore, when considering a head-on collision of equal mass and axial symmetry without spin, let us ponder the null generator of the maximal black room (MBR) which passes through the North Pole, assuming that the axis of axial symmetry aligns with the North-South direction (as depicted in Figure \ref{fig:col}). 
Within this framework, two conceivable scenarios arise for such an MBR generator. The first scenario involves the MBR generator being stationary at the North Pole, while the second scenario entails the generator moving towards the opposite pole in a manner resembling a meridian.

For the first possibility, as the null generator cannot be wandering due to axial symmetry, the generator of the maximal black room (MBR) coincides with the generator of the event horizon. It is worth noting that we exclude the trivial case where the black hole and MBR completely coincide from the following argument.
Assuming that the North Pole lies on the horizon generator and most other points do not, it follows that the null generator approaching the North Pole will take an infinite amount of time to reach it.

In the second scenario, there exist generators of the MBR that resemble meridians at the North Pole. However, in the case of an equal mass black hole collision, at least the equator must correspond to the horizon generator due to the reasons mentioned earlier. Consequently, the meridian generators will not extend to the South Pole and will approach the region of the horizon generator around the equator, which takes an infinite amount of time.

On the other hand, away from the Pole, the asymmetry permits additional courses of the null generator, including an east-west component, which must appear in pairs of eastward and westward null generators due to the spinless nature of the black hole. It is worth noting that this differs from the case of a Kerr black hole\cite{Grover2017}\cite{Grenzebach:2014fha}\cite{Grenzebach2015}\cite{Kobialko2020}, where the presence of spin alters the situation.

\section{geometry of maximal black room}

\subsection{instability of the MBR boundary surface}

In Ref. \cite{Siino2022}, it has been explicated that the MBR is circumscribed by a non-spacelike hypersurface that includes at least one tangential null geodesic at every point. This information has been reiterated in the previous section.
Furthermore, it is convenient to investigate the stability of this boundary surface.
For instance, in a static spherically symmtric spacetime, the circular photon orbit\cite{Virbhadra2000} can be classified into a stable and unstable orbit based on the second derivative of the potential around the extremum.

Considering the MBR of a static spherically symmetric spacetime, it becomes evident that the boundary of the MBR is confined by an unstable circular orbit\cite{Virbhadra2000,Virbhadra2002}, thus implying that the MBR is likely to be bounded by such an unstable boundary. Consequently, to comprehend the black hole shadow in an anisotropic spacetime, it is imperative to analyze the congruence geometry of the rays' surface instead of the photon surface. Nevertheless, the notion of stability is not trivial.

To discuss the unstable nature of the photon orbit around the boundary surface, we consider the congruence of null geodesics\cite{Wald:1984rg} that is orthogonal to the boundary surface. We say that a null geodesic is stable (unstable) if a small deviation is given, and a force for the deviation equation arises that tends to diminish (increase) the deviation, respectively. In this article, we refer to it as locally stable if a restoring force $R_{abcd}l^an^bl^cn^d>0$ arises \cite{Koga2019a}.
Using a similar approach, we will demonstrate the instability of the boundary of the MBR.

In dynamical situations, the concept of local stability may be overly restrictive since local stability and instability can coexist. Therefore, it would be more appropriate to consider the integration of the restoring force over the relevant regions instead of relying solely on the notion of local stability. We define global stability as the integration of the restoring force through the relevant regions, such that the null geodesic has one `conjugate point' along the null geodesic that generates the boundary of the MBR with respect to the congruence that is orthogonal to the boundary surface. The existence of a conjugate point can be anticipated by integrating Raychaudhri's equation, which represents a trace component of the deviation equation.

It is important to note that the presence of a conjugate point does not always indicate that geodesics are reconnecting; it merely implies the vanishing of the Jacobi field. It would be more fitting for our purposes to refer to it as a `reconnection point,' as we aim to exclude the marginal outcomes of the restoring force integration when we establish instability.
We may refer to the location where the conjugate point, accompanied by reconnected geodesics, as the `reconnecting point'.

Furthermore, we must exercise caution when using the term 'globally stable'.  
We  intend to use it in a narrow sense, where the stability implies the existence of another rays' (or photon) surface in neighborhood of the considering boundary surface.
Therefore, a globally stable photon surface may not always be globally stable rays' surface, as a globally unstable photon surface can be stable rays' surface if the surface allows for the presence of neighboring rays' surfaces. The existence of a critical converging surface is an indication of instability, as in the case of the photon sphere.

In this sense, we ought to investigate a subset of the null geodesic congruence, which is initially defined parallel to a null generator of the boundary surface. If a reconnecting point exists for this subset of the congruence within the region being considered, it is equivalent to the existence of another rays' (photon) surface in the vicinity of the boundary surface.

\begin{definition}
Let ${ \cal C}[t]$ denote an initial hypersurface, and let $p$ be a point on its intersection with $\partial {\cal R}$. Consider a smooth family of surfaces generated by null geodesics that are parallel to each other around $p$. For each generating null geodesic, a subset of the null geodesic congruence can be defined in the direction orthogonal to the surface at $p$.
A rays' (photon) surface is globally stable if there exists a reconnecting point for this orthogonal subset of the null geodesic congruence.
\label{def:gst}
\end{definition}

Here, we must exercise caution when discussing the scenario where a photon surface exists in a spacetime without isotropy. This is because the stability for one direction may imply the existence of a rays' surface around the photon surface.
The following proposition is therefore limited to cases where the boundary of the maximal black room is not a photon surface.

\begin{proposition}
The boundary of the maximal black hole region (MBR), which is not a photon surface, lacks global stability in the direction orthogonal to the boundary surface.
\label{prop:gst}
\end{proposition}

{\it Proof:}
Assuming global stability of $\partial{\cal R}$ in the orthogonal direction, the reconnecting point arises by definition\ref{def:gst}. 
 We have excluded the scenario where the boundary surface is a photon surface everywhere from our discussion. Regarding global stability, if the surface is not a photon surface, it is sufficient to focus only on the case where null geodesics enclosed in the boundary are aligned in a particular direction, in general, to demonstrate that a stable configuration is not possible. Additionally, we can consider the MBR denoted by $\cal R$ and $\partial {\cal R}$, which is a smooth boundary around the point $p\in \partial {\cal R}$. Then, from proposition 2 of Ref.\cite{Siino2022}, there exists $l_p$, which is a null geodesic through $p$ contained in $\partial {\cal R}$.

Based on the definition\ref{def:gst}, a suitable selection of ${\cal C_0}[t]$ results in the family of surfaces in definition \ref{def:gst} being orthogonal to ${\cal C}[t]$, as the boundary surface is non-spacelike.
Consider an initial hypersurface ${\cal C_0}[t_0]$ and a point $p\in \partial{\cal R}\cap{\cal C_0}[t_0]$ on it.
For the boundary of the MBR $\partial{\cal R}$, a generating null geodesic $l_p\in \partial{\cal R}$ passes through a point $p$. We can then define a deviation vector $v^a(t)$ around $l_p$ such that it is initially orthogonal to $\partial{\cal R}$. As a result, the subset of the null geodesic congruence defined by $v^a$ becomes hypersurface orthogonal. 
Subsequently, we shall examine the subset of the null geodesic congruence defined by a deviation vector $v^a$ which satisfies $v^a(t_0)\perp \partial {\cal R}$, and is thus hypersurface orthogonal.

If $l_p$ has a reconnecting point, we need to consider three possible types of global stability. The first type is stability within the surface, where another null geodesic $l_p'$ exists from $p_i\in \partial{\cal R}$ to $p_f\in \partial{\cal R}$, lying within the closure $\overline{\cal R}$ of $\cal R$. The second type is stability outside the surface, where $l_p'$ never lies inside the interior of $\cal R$. The third type is stability on both sides of the surface.

When considering the stability of the region within the boundary surface, where the null geodesic $l_p'$ lies in the interior of $\cal R$, it becomes clear that a null geodesic will exit $\cal R$ near $l_p'$ after entering it. Therefore, the presence of such a null geodesic $l_p'$ within the surface immediately rules out the possibility of stability inside or on both sides of the black room.

The prevailing notion is that the boundary does not constitute an entirely photon surface, but rather rays' surface where only several tangential null geodesics $l_p$'s lies on $\partial {\cal R}$. In this case, stability outside the surface indicates that the black room is not maximal, since the possibility of an entirely photon surface has been ruled out.

Now we shall consider the stability outside of the surface in the null direction of $l_p$ around $p$.
According to the definition of stability, for a sufficiently small ${\cal U}_p$, there is a point $p^{\prime\prime}$
outside of $\overline{\cal R}$ that produces null geodesics $l_{p^{\prime\prime}}$ through $p^{\prime\prime}$
that are close to $l_p$ along $\partial\cal R$ at $p$ (in the topology Eq.(3.1) of Ref. \cite{Siino2022}). Subsequently, $l_{p^{\prime\prime}}$ intersects $\partial\cal R$ on both sides of $p^{\prime\prime}$ on ${\cal U}p$. 
As a result, there exists a small deformation ${\cal R}^\prime\supset \cal R$ of $\cal R$ in ${\cal U}p$ such that $\cal R'$ has the $l{p^{\prime\prime}}\cap {\cal R}^c$ on its boundary $\partial \cal R'$ without having two intersections with the null geodesics $l'_{p^{\prime\prime}}$ which is through $p^{\prime\prime}$ in other direction of $l_p^{\prime\prime}$.
Of course, ${\cal U}_p$ can include no null geodesic escaping $\cal R^\prime$ after entering $\cal R^\prime$.
Moreover there is such a small ${\cal U}_p$ that $l_{p^{\prime\prime}}$ never enters $\cal R$ even outside of ${\cal
U}_p$ since $\cal R$ is open subset.
Then the stability outside of the surface is not allowed so as not to give another black room which is not subset of $\cal R$.

\qed\\

{\bf Remark}
{\it For the case of an entirely photon surface without isotropy, for the stability outside the surface, the discussion is not in question, but the stability inside the MBR is consistently absent.
When one direction is stable inside the surface, there is the possibility that small deformations along $l_p'$ cannot help but produce invading null geodesic near $l_p''$ which is located very close to the null generator in this direction.

However, with a restriction to the discussion of isotropic situation, the argument changes such that
for the stability outside the surface, all direction theres will possess $l_p'$, and this implies that symmetric deformations result in a larger black room, considering the symmetry. This is applicable for the spacetime that are fully symmetric, such as a photon sphere.
}

In the following section, we will see an example that the boundary surface exist through the locally stable and unstable regions.



\subsection{black room evolution}

Now, we are addressed to examine how to determine the MBR.

Although the mathematical description of the MBR has been provided in Ref.\cite{Siino2022}, it is not unique in the case of spacetime with a given initial hypersurface.
Even in the context of Schwarzschild spacetime, a sequence of photon surfaces can be recognized as the boundary of a MBR as it is bounded in the case with an initial hypersurface, despite being unbounded within the overall spacetime, as indicated below.
Rather, we should impose conditions based on astrophysical considerations, as the existence of the initial hypersurface would not be free from the moment the photon source begins to emit light.
Simultaneously specifying both the initial hypersurface and boundary conditions is imperative in determining a scientifically rigorous boundary condition with astrophysical implications, ensuring that information degeneration is avoided.

 \begin{figure}[hbtp]
\begin{center}
\includegraphics[height=12cm]{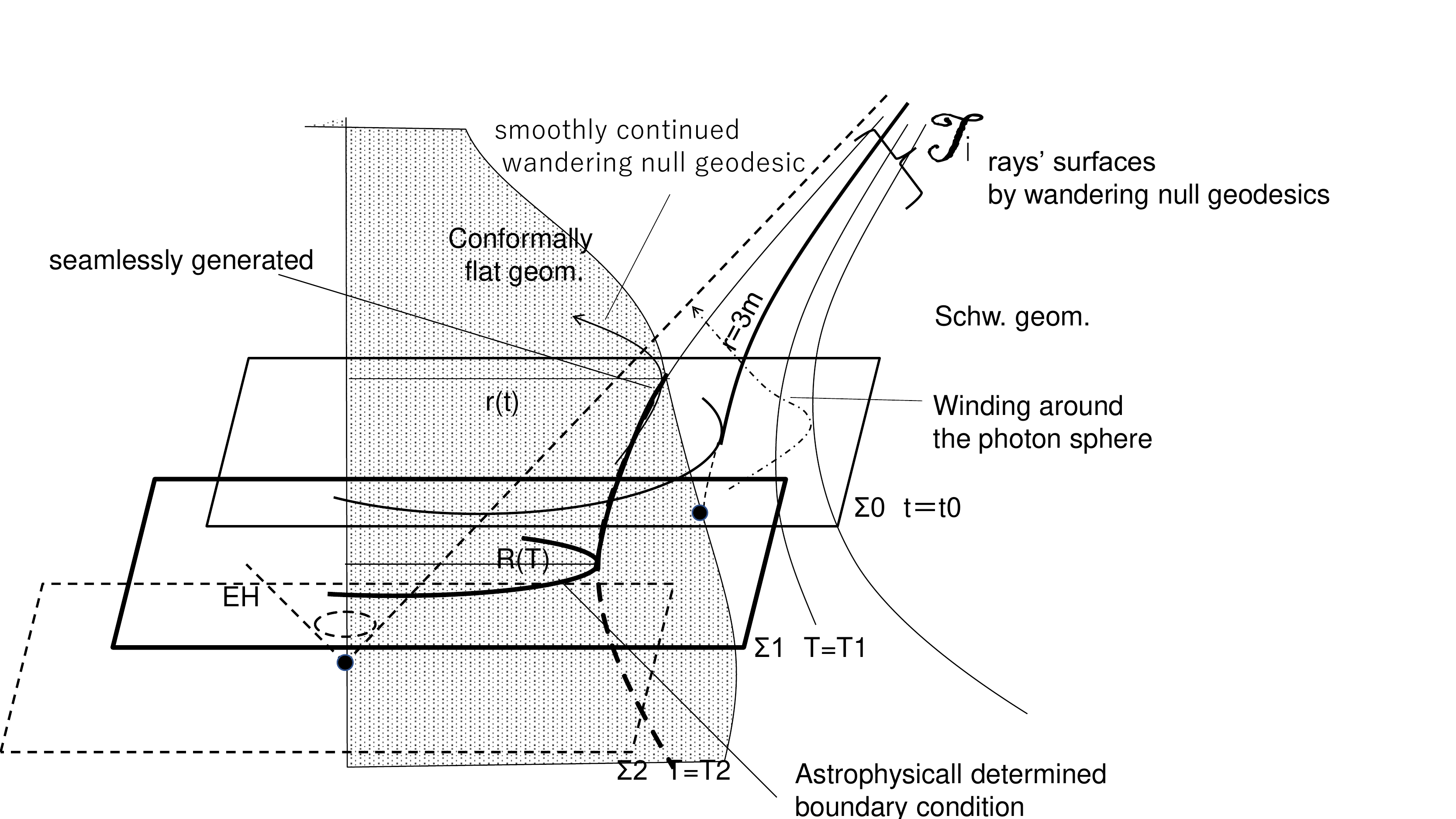}
\caption{A homogeneous dust star with spherical symmetry is undergoing gravitational collapse. Outside the star, the spacetime is characterized by the Schwarzschild metric, and there are a family of rays' surface ${\cal T}_i$ that correspond to timelike hypersurfaces generated by rotating a null geodesic with respect to $SO(3)$ symmetry. These rays converge to an unstable circular orbit.
On the other hand, inside the star, where the geometry is conformally flat, the rays' surface can be described by the one-sheeted hyperboloid with a throat located on the hyperplane $T=T_1$, that is centered around the conformally flat coordinates $(T=T_1,X=0,Y=0,Z=0)$.}
 \label{fig:evo}
 \end{center}
\end{figure}

It is widely acknowledged that the spherically symmetric timelike hypersurface on which future wandering null geodesics lie is not unique even in the Schwarzschild spacetime. 
\begin{align}
{\rm d}s^2=-(1-2m/r){\rm d}t^2+\frac1{1-2m/r}{\rm d}r^2+r^2({\rm d}\theta^2+\sin^2\theta{\rm d}\phi^2).
\label{eq:schw}
\end{align}
Any point outside the event horizon will cast future null geodesics with a specific impact parameter that infinitely winds around the photon sphere\cite{Misner1973}.

The equation governing the radial motion of the effective potential is given by:
\begin{align}
\left(\frac{dr}{d\lambda}\right)^2+B^{-2}(r)=b^{-2},\\
B^{-2}(r)=r^{-2}(1-2m/r),\\
b=({\rm impact}\ {\rm parameter}).
\end{align}
The supplementary equations for determining the angular and time motion are as follows:
\begin{align}
\left(\frac{d\phi}{d\lambda}\right)&=1/r^2,\\
\left(\frac{dt}{d\lambda}\right)&=b^{-1}(1-2m/r)^{-1}.
\end{align}

From the shape of the effective potential $B^{-2}(r)$, we derive the following conclusions:
\begin{enumerate} 
\item 
A massless particle with zero rest mass, having an impact parameter $b$ greater than $3\sqrt{3}m$ and approaching from radial infinity $r=\infty$, undergoes reflection off the potential barrier at periastron $b=B$, where $dr/d\lambda=0$, and subsequently travels back to infinity.
\begin{itemize}
\item If $b\gg 3\sqrt{3}m$, the orbit approximates an almost straight line with a deflection angle of $4m/b$.
\item For $0<b-3\sqrt{3}m\ll m$, the particle circles the star many times (unstable circular orbit) at $r\sim m$ before flying back to $r=\infty$.
\end{itemize}
\item A massless particle with zero rest mass, with $b<3\sqrt{3}m$ falling in from $r=\infty$, falls into $r=2M$ (no periastron).
\item A massless particle with zero rest mass, emitted from near $r=2m$, escapes to infinity only if it has $b<3\sqrt{3}m$; otherwise, it reaches an apastron and then gets pulled back into $r=2m$.
\end{enumerate}
We notice that the impact parameter $b$ is reinterpreted for scattering from inside the photon sphere $r=3m$.

As previously indicated, for a general black hole, the boundary of the MBR should not be considered globally stable (Prop. \ref{prop:gst}). Consequently, it is expected that a specific impact parameter could be obtained for a range of initial conditions of null geodesics. Hence, in the context of an eternal black hole, the requirement of spatial delimitation for the MBR in past infinity would allow for the determination of the MBR in the manifold with $t\in (-\infty,\infty)$. The future wandering null geodesic must be a totally wandering null geodesic, and the rays' surface of the MBR boundary can be uniquely determined for this eternal black hole.

Contrarily, in the manifold of $t\in (t_0,\infty)$, the situation is less straightforward. For an asymptotic boundary condition at future timelike infinity, the global instability of the boundary of the MBR as a rays' surface permits the determination of a unique boundary surface through a proper combination of the choice of initial hypersurface and initial data.

As a suitable initial condition, we put forth the following suggestions:
\begin{itemize}
\item The initial hypersurface should align with the inception of the relevant photon sources.
\item The trajectory of photons can be established by imposing an orthogonal boundary condition, whereby the boundary of the MBR is perpendicular to the initial hypersurface.
\end{itemize}

Now, we shall demonstrate the discovery of a time-dependent MBR within a dynamically evolving spacetime. Submanifolds of a spherical, homogeneous, collapsing dust star spacetime\cite{Oppenheimer:1939ue}\cite{Cao2021} will be analyzed in the following analysis.

In the context of a collapsing homogeneous spherical star (reference \cite{Oppenheimer:1939ue}), we observe the existence of a timelike hypersurface $r=3m$ that corresponds to the photon sphere of the Schwarzschild spacetime outside of the star. If the spacetime manifold under consideration, denoted as $M=\{\Sigma(t)|t>t_0\}$, does not encompass the intersection of the timelike hypersurface with the surface of the star, it is anticipated that the boundary of the MBR is congruent to the said surface located at $r=3m$.

From the proposition 2 in Ref.\cite{Siino2022}, it can be deduced that the null geodesic situated on the boundary is a wandering null geodesic $\lambda_w$. If its starting point, represented by $p$, is given as $p=(t_0,r_0,\theta_0,\phi_0)$ in a spherical coordinate, then the null geodesic $\lambda_w$ can be characterized as a spiraling solution of the null geodesic equation with initial data $\lambda_w[t_0,r_0,\theta_0,\phi_0](\tau=0)=p(t_0,r_0,\theta_0,\phi_0)$, which winds around the unstable photon sphere located at $r=3m$. 

The spherical symmetry gives rise to a timelike hypersurface of the rays' surface ${\cal T}_i$, which is generated by the wandering null geodesic $\lambda_w(\tau)$, defined as
\[{\cal T}_i=\bigcup_{0\leq\theta< \pi,0\leq\phi<2\pi} {\lambda_w[t_i,r_i,\theta,\phi](\tau)}.\] 
As per the demonstration provided in theorem 3 of Ref.\cite{Siino2022}, the exterior of the horizon is foliated by the family of these surfaces ${{\cal T}_i}$.
Subsequently, a rays' surface should be selected from among them as the boundary of the MBR, based on physical requirements.
In this particular scenario, the preferred choice for the initial surface would be the Schwarzschild timeslicing, accompanied by the orthogonal condition serving as the boundary condition. This choice is considered a natural and appropriate means for determining the trajectory of photons. Consequently, the rays' surface at $r=3m$ is selected, aligning with our desired outcome.

We now consider a scenario in which the spacetime manifold $M={\Sigma(T)|T>T_1}$ encompasses the past edge of the photon sphere at $r=3m$ on the surface of the star, as shown in a simultaneous plane $\Sigma_1=\Sigma(T=T_1)$ in Fig. \ref{fig:evo}. We make use of two time coordinates $t$ and $T$ for convenience.

Within the confines of the homogeneous star, the geometry described by the conformally flat metric can be expressed as:
\begin{align}
{\rm d}s^2&=\Omega(T,R)^2(-{\rm d}T^2+{\rm d}X^2+{\rm d}X^2+{\rm d}Z^2) \\
&=\Omega(T,R)^2(-{\rm d}T^2+{\rm d}R^2+R^2({\rm d}\theta^2+\sin^2\theta{\rm d}\phi^2)).
\label{eq:star}
\end{align}
The metric dictates that the boundary surface of the MBR is defined as a one-sheet hyperboloid  with a throat located on the hyperplane $T=T_1$, as the null geodesic is conformally mapped to itself.  This hyperboloid acts as a photon surface that is perpendicular to the initial hyperplane $T=T_1$ in the Minkovski spacetime$M^{(1,3)}$, as mathematically expressed by the equation:
\[-(T-T_1)^2+X^2+Y^2+Z^2=l(T_1)^2, \]
where $l(T_1)$ will be determined for the initial time $T_1$ through the smooth continuation at the surface of the homogeneous star.
The one-sheet hyperboloid is generated by the O(3) rotational isometry, denoted as ${\cal S}=\{ T(\gamma_0) | {}^\forall T\in \rm{O(3)} \}$, where
$\gamma_0$ represents an arbitrary null geodesic in $M^{(1,3)}$.

As the angular coordinates are shared by Eq. \ref{eq:schw} and Eq. \ref{eq:star}, the rays' surface can be smoothly joined at the surface of the homogeneous star, where the null geodesic remains smoothly continuous.
Under these circumstances, the boundary of the MBR should also be selected from the surface ${\cal T}_i$, and the generating wandering null geodesic must be continued seamlessly at the surface of the homogeneous star.

In this specific study, the proposal is to opt for a constant value of $T$ as a trial for the initial hypersurface, while maintaining the validity of the orthogonal condition. This approach results in a well-defined boundary for the MBR.
Our initial radius of the MBR, denoted as $R(T_1)$, will be set to $l(T_1)$ at the moment $T=T_1$, ensuring that the timelike hypersurface of rays on the one-sheet hyperboloid is perpendicular to the initial hyperplane defined by $T=T_1$, while maintaining smooth continuation at the surface of the homogeneous star.

Furthermore, it is noteworthy that at the formation time $T=T_2$ of the event horizon, as depicted in Figure \ref{fig:evo}, the boundaries of the MBR converge towards the event horizon, with the one-sheet hyperboloids tending towards a cone described by the equation $-(T-T_2)^2+X^2+Y^2+Z^2=0$. This observation is in agreement with the fact that the MBR coincides with the black hole when taking into account the formation of the event horizon. This mathematical evidence provides support for the consistency of our proposal.

Subsequently, we shall gain insight into how the black shadow encapsulates crucial chronological information, thereby highlighting the need for prudence when deducing information about the black hole shadow solely from concurrent measurements, such as the ADM mass of the black hole.

Presently, we redirect our focus to the overarching problem of determining the MBR in relation to a globally unstable wandering null geodesic. In general circumstances, the existence and uniqueness of the solution for the null geodesic are contingent upon the alignment of the null geodesic congruence with conjugate points. The potential for global instability and accumulation around the wandering null geodesic congruence was partially explored in \cite{Siino2020}, and, for instance, may be guaranteed by asymptotic flatness. Moreover, in a more general situation, with the instability of the solution, a definitive solution with normal direction dependence can be procured, regardless of whether it is derived analytically or computationally.

Suppose that the MBR under the demanded efficacy exists.
In the distant future, an infinite number of conjugate points can be found on the boundary hypersurface of the generating future wandering null geodesic. However, none seems to exist in the direction vertical to the boundary hypersurface, indicating global instability.   The absence of reconnecting point for global instability (as defined in Definition \ref{def:gst}) implies that a wandering null geodesic$\lambda_p$ from a point $p$ may  uniquely exist in order to determine the boundary hypersurface.
Caution must be exercised in this context, as the absence of a reconnecting point has not yet been definitively proven to guarantee the unique existence of a wandering null geodesic $\lambda_p$ for generating the boundary, as documented in \cite{Siino2022}.

When the globally unstable rays' surface with the sufficient efficacy, which is the timelike boundary of the MBR, exists, it is expected that the conjugate point of the generators will exist in a subset of the congruence in the tangential direction for consistency. 
Locally acknowledging the analogy with the aforementioned spherically symmetric spacetime, one might consider a local criterion to select the boundary from among the family of locally parallel rays' surfaces by the orthogonal condition.
Nonetheless, there remains an ambiguity in the direction of the generator of the boundary hypersurface along the $S^1$ direction.

To address this, we can identify a point $p$  that lies on both the timelike boundary hypersurface and the spatial initial hypersurface. Subsequently, we can specify the available initial conditions of the wandering null geodesic from $p$ in such a way that it generates the boundary hypersurface along the $S^1$ direction (as seen from the proof of the theorem 3 in \cite{Siino2022}), while ensuring that the timelike boundary hypersurface is perpendicular to the initial spacelike hypersurface. Then, assuming a direction for the generator, we can locally determine a 2-surface $\cal B$ in the neighborhood of $p$, on the initial hypersurface as a section of the boundary hypersurface. 
Hence, we anticipate that a particular integration technique could provide a comprehensive solution, resulting in a closed 2-dimensional surface on the initial hypersurface, provided that we choose a direction that makes the 2-surface closed, although achieving this in a general sense could pose a challenge.

When axial symmetry is present\footnote{It would be appeared in our forthcoming work.}, the direction of the generator can be determined by the orthogonal condition, and the condition for $\cal B$ to be closed reduces to a one-dimensional determination. With the aid of appropriate integration techniques, these conditions will enable the selection of $\cal B$ from the set of rays' surfaces, analogous to the shooting method used to identify the apparent horizon in an axially symmetric spacetime\cite{Nakamura1984}\cite{Oohara1985}.

Additionally, it is worth mentioning that in a general context, the globally unstable rays' surface may not be uniquely determined, and alternative solutions may exist in entirely different directions. For example, in the context of spherically symmetric static spacetime, one can effortlessly generate instances by manipulating the metric function to create a potential function with multiple minima and maxima for the orbiting null geodesic equation, where both stable and unstable circular orbits coexist. In such cases, it is essential to discard the stable surface promptly, and select the outermost unstable one manually.

Upon completion of the analysis in general circumstances, all of these aspects will be considered, and it will reveal the total nature of the MBR, akin to the previously mentioned spherically symmetric case.

\section{black room and rays' surface black hole}

Here, we delve into the subject of the emergence of the MBR. The discourse is comprised of two focal points: one pertains to its existence, the other to its correlation with an exotic celestial body.
Before delving into an investigation of the existence of the MBR and its connection to the rays' surface, it is imperative to refresh our memory regarding the established knowledge regarding the photon surface in a static, spherically symmetrical spacetime, as the photon surface serves as a special case of the rays' surface.

In the context of a static and spherically symmetric spacetime, it is well established that the boundary of the MBR manifests as an unstable extremal of the potential function for a null geodesic equation.
In a static, spherically symmetrical spacetime, the geometry of the black hole shadow is depicted by a photon sphere. As such, there is a wealth of knowledge regarding the relationship between the photon sphere and compact celestial objects.
For the purpose of comparison, we will re-emphasize the correlation between the photon sphere and photon surface in a spherically symmetric spacetime, and thus, deduce that the presence of a photon surface implies the existence of a black hole spacetime.

Following is the known fact.
\begin{theorem}[Claudel, Virbhadra, Ellis]
It was shown that, in spherically symmetric spacetime, a black hole is surrounded by the photon sphere under an energy condtion.
It was shown that, in spherically symmetric spacetime, any photon sphere surround a black hole, a naked
singularity or more than a certain amount of matter.
\label{thm:CVE}
\end{theorem}

We shall now investigate these aspects in the context of the time-dependent boundary hypersurface of the MBR. This shall be denoted as the general manifestation of the MBR for a black hole, and as demonstrated in the preceding section, we have established its existence in the Oppenheimer-Snyder spacetime.

\subsection{in a black hole spacetime, there is a MBR.}

Despite the prevailing notion that no light signal can escape a black hole, we contemplate the possibility that a photon originating from within the region of the maximal black room (MBR) could potentially possess the ability to evade capture.
The domain of the black hole serves as a trivial black room, despite its absence of an escape route.
Hence, it is conceivable that numerous black rooms could simultaneously exist within a single black hole spacetime. The existence of a maximum black room among them would therefore bestow universal significance in the optical investigation of black holes, notwithstanding the inevitable perturbation resulting from the specific configuration of light sources.

Through the principles of deductive reasoning, it is apparent that an MBR endures uniformly in arbitrary black hole spacetimes. 
As per definition, the region occupied by the black hole constitutes the black room $R_1$. Subsequently, if an additional black room $R_2$ exists that is not encompassed within $R_1$, then a third black room, $R_3 = R_1 \cup R_2$, is derived.
Through this iterative process, an MBR can be inferred via Zorn's lemma\cite{Zorn1935}, provided that its volume is finite.
\begin{theorem}
In asymptotically flat spacetime\footnote{For example, in a compact universe the statement will not hold.}\cite{Hawking:1973uf,Wald:1984rg},
any black hole region is contained by the maximal black room and is surrounded by the rays' surface.
\label{th:2}
\end{theorem}
{\bf Proof:}
As per its definition, the black hole is tantamount to a black room. 
If it does not comprise the MBR, its union will form another black room denoted as $R_3 = R_1 \cup R_2$.

Zorn's lemma\cite{Zorn1935}, a fundamental result in mathematics, states that any partially ordered set that contains upper bounds for every chain must contain a maximal element.
Given a set, its power set (the set of all its subsets) ordered by inclusion forms a partially ordered set, and hence Zorn's lemma applies. This implies the existence of a maximal subset $R_M$ in the power set ${R_i}$.

Conversely, there is an upper bound of the size of MBR even if the maximal subset $R_M$ was unbounded.
In an asymptotically flat spacetime, it is impossible for a spatially bounded black room of appropriate size denoted by $R_M'$ to be contained within a larger black room.

By the definition of asymptotic flatness, the complement of the subset $R_M'$ in the spacetime, denoted as $M''=M\setminus R_M'$, exhibits asymptotic simplicity, and its causal structure tends towards that of Minkowski spacetime. In other words, there exists no closed rays' surface that is spatially bounded for any arbitrary spatial section in Minkowski spacetim.
In a nearly Minkowski spacetime, all rays' surfaces approach future null infinity, as its generating null geodesic cannot be a wandering null geodesic due to the asymptotic simplicity of the spacetime.
\qed

Therefore the black hole region is surrounded by the rays' surface.
Then the ensuing Corollary amplifies the potency of the theorem \ref{thm:CVE}.
\begin{corollary}
Considering a spherically symmetric spacetime.
The black hole is surrounded by a spherical photon surface.
\end{corollary}
Even if the energy condition is not satisfied, the theorem\ref{thm:CVE} remains partially valid and ensures that the surface of the rays becomes a photon surface, specifically under spherical symmetry.
 In the static scenario, the spherical photon surface assumes the role of the photon sphere.

\subsection{exotic stars with photon sphere in spherically symmetric spacetime}

In spite of theorem \ref{th:2}, it is possible for an unstable photon sphere to exist in spherically symmetric spacetime without an event horizon. In this article, the term `exotic star' denotes a celestial body that possesses a photon sphere but is not a black hole. Examples of such bodies include gravastars\cite{Mazur2023} or regular black holes\cite{Fan2016}\cite{Stuchlik2019} that lack an ordinary matter field.

If an MBR is enclosed by an outgoing null surface, it necessarily corresponds to an event horizon, as it is a bounded region with no null geodesic escaping. Let us now consider a spherically symmetric exotic star whose MBR is enclosed by a timelike hypersurface.

For an MBR bounded by a timelike hypersurface, it has both ingoing and outgoing null geodesics on the radial direction at a point on its boundary, entering the interior of the MBR and escaping towards ${\cal I}^+$, respectively. Such escaping null geodesics may be referred to as null lines\cite{Galloway2000}.
A null line in a spacetime $(M, g)$ is an inextendible null geodesic that is globally achronal, signifying that it is impossible to connect any two points on the geodesic by means of a timelike curve. In this article, we also use the term `null line' even for future directed null geodesic with an starting point.
Of course, the existence of null lines is dependent on the presence of conjugate points, and some important results were derived by Galloway\cite{Galloway2002}.

For any given point $p$ located within the interior of the MBR, every null geodesic emanating from $p$ can either be classified as a wandering null geodesic which may approach the timelike infinity $i^+$\cite{penrose1986spinors}\cite{Geroch:1972un}, or alternatively as partially a null line that reaches ${\cal I}^+$ beyond a certain designated point $p'$ on its trajectory. However, due to the spherical symmetry, it can be inferred that the null geodesic in the radial direction cannot be a wandering null geodesic. In a spherically symmetric spacetime, a radial outgoing null geodesic can only have conjugate points if an apparent horizon is formed.
Assuming the null geodesic is geodesically complete, it will propagate towards ${\cal I}^+$ without escaping from the MBR.

Hence, it can be inferred that either geodesic incompleteness or the presence of a child universe with another ${\cal I}^+$ is a feasible proposition. Therefore, within the confines of such an MBR, it is impossible for an exotic star to exhibit the causal structure of a standard star.

In this discussion, we aim to investigate the material origins underlying the formation of a child universe. A straightforward computation utilizing the spherical metric expressed as
\begin{align}ds^2&=-dt^2+X^2(r,t)dr^2+A^2(r,t)(d\theta^2+sin^2\theta d\phi^2) \\
&=-dt^2+\frac{A'^2}{1-k(r)}dr^2+A^2(r,t)(d\theta^2+\sin^2\theta d\phi^2)
\end{align}
yields a definite expression of the Einstein equation, which can be written as
\begin{align}
\frac23\frac{\ddot{A}}{A}+\frac13\frac{\ddot{A}'}{A'}=-\frac{4\pi G}3(\rho+3P),
\end{align}
 where the Hamiltonian constraint is utilized when dealing with a perfect fluid.  It is shown that even if the strong energy condition holds, i.e., $\rho+3p \geq 0$, the radial expansion may lead to an acceleration in the deceleration $A'$ in the angular direction \cite{Enqvist2007}. 
Therefore, the scenario where the deceleration $A'$ in the angular direction dominates corresponds to a void universe or a child universe, which violates the strong energy condition. This case might correspond to the scenario where the theorem in \cite{Claudel:2000yi} holds for a large amount of matter.


Thus the discussion leads to the conclusion that there exists an event horizon or a naked singularity or child universe. 
Here it should be added that a similar discussion can be made in the axially symmetric situation about the null geodesics on the z-axis.

Hence, the analysis above implies that there exists an event horizon, a naked singularity, or a child universe. It is worth noting that a similar discourse can be conducted with respect to null geodesics along the z-axis in the situation of axial symmetry.

In the preceding discussion, the presence of spherical symmetry (or, in some cases, axial symmetry) was crucial in precluding the possibility of a wandering null geodesic towards the central region. However, if we assume that the center of the star is not a naked singularity from a geometrical standpoint, at least for small perturbations, we can anticipate a lack of significant alteration in the trajectories of null geodesics. Consequently, we can expect that this reasoning holds true for a broad range of nearly symmetric situations, and any perturbative analysis would likely serve to elucidate the matter.


Undoubtedly, even if one detects a deviation from the literal definition of the MBR\cite{Wielgus2020}, it does not firmly establish the existence of an exotic star, as it could be obscured by a light source located in its vicinity\cite{Keir2016,Cardoso2014,Cunha2017a}. Nevertheless, with the advancement of technology and increasing scientific understanding, it may become feasible to distinguish unique features, such as distinctive patterns in the data captured by imaging or proof of interactions with neighboring matter, thus enabling differentiation between the exotic star and conventional black holes.







\section{summary}

The significance of the maximal black room (MBR) has undergone scrutiny in our research. Our findings indicate that the MBR is inherently globally unstable, being a natural generalization of the photon sphere. We established the MBR by considering its initial hypersurface and conditions, particularly in the context of spherical homogeneous dust collapse\cite{Oppenheimer:1939ue}\cite{Cao2021}. Additionally, we have explored the possibility of using the MBR to distinguish between traditional black holes and exotic celestial bodies, such as the gravo-star or regular black hole.

Examining exotic stars would not meet the required level of astrophysical validity, particularly in light of their postulated transparency.

 \bibliographystyle{unsrt}
\bibliography{whiteroom}

\end{document}